\documentclass{ifacconf}

\usepackage{amsmath}
\usepackage{IEEEtrantools}
\usepackage{mathtools}
\usepackage{./TLC}
\usepackage{blox}
\usepackage{physics}
\usepackage{graphicx}      
\usepackage{natbib}        
\begin{document}
\begin{frontmatter}

\title{Loop Shaping with\\Scaled Relative Graphs} 

\thanks[footnoteinfo]{The research leading to these results has received funding from the European
Research Council under the Advanced ERC Grant Agreement Switchlet n. 670645, and from
the Cambridge Philosophical Society.}

\author[First]{Thomas Chaffey} 
\author[First]{Fulvio Forni} 
\author[First]{Rodolphe Sepulchre}

\address[First]{University of Cambridge, Department of Engineering, Trumpington Street,
Cambridge CB2 1PZ, {\tt\small \{tlc37, ff286, rs771\}@cam.ac.uk}.}

\begin{abstract}                
        The Scaled Relative Graph (SRG) is a generalization of the Nyquist diagram that may
        be plotted for nonlinear operators, and allows nonlinear robustness margins
        to be defined graphically.  This abstract explores techniques for
        shaping the SRG of an operator in order to maximize these robustness margins.
\end{abstract}


\end{frontmatter}
\section{Introduction}

Loop shaping is one of the earliest methods of controller design, originating in the
work of Nyquist, Bode, Nichols and Horowitz on feedback amplifiers \citep{Bode1960}.
The basic principle of loop shaping is to tune a system's closed loop performance by adjusting the open loop frequency response. Robustness is
captured by the distance of the Nyquist diagram from the point $-1$; closed loop performance is captured by the
 sensitivity and related transfer functions. 
Loop shaping is still widely used in industry today -- the graphical nature of the
tool gives a clear view of the design tradeoffs between performance and
robustness.  Even in the age of modern optimal and robust control, loop shaping remains
a core tool for the control engineer.
The idea of enlarging stability margins eventually led to the Zames' formulation of
$H_\infty$ control \citep{Zames1981a},
 and some of the most successful methods of
robust control combine $H_\infty$ control with classical loop shaping
ideas \citep{Vinnicombe2000, McFarlane1992}.

The Scaled Relative Graph (SRG) is a graphical representation of a nonlinear operator, recently introduced in the theory
of optimization by \cite{Ryu2021}. The SRG allows simple, intuitive
proofs of convergence for optimization algorithms, and allows optimal convergence
rates to be visualized as distances on a plot.  The authors have recently connected the SRG to classical
control theory, showing that it generalizes the Nyquist diagram of an LTI transfer
function \citep{Chaffey2021c}.  A range of incremental stability results, including
the Nyquist and circle criteria, small gain and passivity theorems and secant
condition, can be interpreted as guaranteeing the separation of the SRGs of two
systems in feedback, and
this interpretation has led to new conditions for incremental stability
\citep{Chaffey2021d}.  The distance between the two SRGs is an incremental disc
margin, the reciprocal of which bounds the incremental gain of the closed loop.  
The SRG makes the design intuition afforded by the Nyquist diagram
available for nonlinear systems.

This abstract describes ongoing research into the use of SRGs for loop shaping nonlinear feedback systems. 
It
has long been observed that introducing nonlinearity can overcome fundamental
limitations of LTI control 
-- for example, the describing function of the Clegg integrator has a phase lag of
only 38$^\circ$, rather than the usual $90^\circ$ of a linear integrator
\citep{Clegg1958}.  This
motivates a better understanding of how nonlinearities may be used to shape the
performance of a feedback system.

\section{Review of Scaled Relative Graphs}

We begin this extended abstract with a brief review of the theory of SRGs.  

\subsection{Signal Spaces}
We describe systems using operators, possibly multi-valued, on a Hilbert space.
A Hilbert space $\mathcal{H}$ is a vector space equipped with an inner product,
$\bra{\cdot}\ket{\cdot}: \mathcal{H} \times \mathcal{H} \to \mathbb{C}$, and the
induced norm $\norm{x} \coloneqq \sqrt{\bra{x}\ket{x}}$.

We will pay particular attention to the Lebesgue space $L_2$. 
Given  $\mathbb{F} \in \{\R, \C\}$, $L^n_2(\mathbb{F})$ is defined as the set of
signals $u: \R_{\geq 0} \to \mathbb{F}^n$ such that
\begin{IEEEeqnarray*}{rCl}
        \norm{u} \coloneqq \left(\int_{0}^{\infty} u(t) \bar{u} (t) \dd{t}\right)^{\frac{1}{2}}
        < \infty,
\end{IEEEeqnarray*}
where $\bar{u}(t)$ denotes the conjugate transpose of $u(t)$.
The inner product of $u, y \in L_2^n(\mathbb{F})$ is defined by
\begin{IEEEeqnarray*}{rCl}
        \bra{u}\ket{y} \coloneqq \int_{0}^{\infty} u(t) \bar{y} (t) \dd{t}.
\end{IEEEeqnarray*}
The Fourier transform of $u \in L_2^n(\mathbb{F})$ is defined as
\begin{IEEEeqnarray*}{rCl}
        \hat{u}(j\omega) \coloneqq \int_0^\infty e^{-j\omega t}u(t) \dd{t}.
\end{IEEEeqnarray*}
We omit the dimension and field when they are immaterial or clear from context.

\subsection{Relations}

An \emph{operator}, or \emph{system}, on a space $\mathcal{H}$, 
is a possibly
multi-valued map $R: \mathcal{H} \to \mathcal{H}$.
The identity operator, which maps $u \in \mathcal{H}$ to itself, is denoted by $I$.
The \emph{graph}, or \emph{relation}, of an operator, is the set $\{u, y\; | \; u \in
\dom{R}, y \in R(u)\} \subseteq \mathcal{H}\times\mathcal{H}$.  We use the notions of an operator and its relation
interchangeably, and denote them in the same way.

The usual operations on functions can be extended to relations.  Let $R$ and $S$ be
relations on an arbitrary Hilbert space.  Then:
\begin{IEEEeqnarray*}{rCl}
        S^{-1} &=& \{ (y, u) \; | \; y \in S(u) \}\\
        S + R &=& \{ (x, y + z) \; | \; (x, y) \in S, (x, z) \in R \}\\
        SR &=& \{ (x, z) \; | \; \exists\; y \text{ s.t. } (x, y) \in R, (y, z) \in S \}.
\end{IEEEeqnarray*}
Note that $S^{-1}$ always exists, but is not an inverse in the usual sense.  In
particular, it is in general not the case that $S^{-1}S = I$.  The relational inverse
plays a fundamental role in the techniques described in this abstract.  Rather than
directly shape the performance of a negative feedback interconnection, we will shape
the performance of its inverse relation -- a parallel interconnection.

These operations will also be used on sets of operators, with the meaning that the
operations are applied elementwise to the sets (under the implicit assumption that
the operators have compatible domains and codomains).

\subsection{Scaled Relative Graphs}

We define SRGs in the same way as \cite{Ryu2021}, with the minor modification of
allowing complex valued inner products. 

Let $\mathcal{H}$ be a Hilbert space.
The angle between $u, y \in \mathcal{H}$ is defined as
\begin{IEEEeqnarray*}{rCl}
        \angle(u, y) \coloneqq \acos \frac{\Re \bra{u}\ket{y}}{\norm{u}\norm{y}}. 
\end{IEEEeqnarray*}

Let $R: \mathcal{H} \to \mathcal{H}$ be an operator.  Given $u_1, u_2 \in
\mathcal{U} \subseteq \mathcal{H}$, $u_1 \neq u_2$, define the set of complex numbers $z_R(u_1, u_2)$ by
\begin{IEEEeqnarray*}{rCl}
        z_R(u_1, u_2) \coloneqq &&\left\{\frac{\norm{y_1 - y_2}}{\norm{u_1 - u_2}} e^{\pm j\angle(u_1 -
u_2, y_1 - y_2)}\right.\\&&\bigg|\; y_1 \in R(u_1), y_2 \in R(u_2) \bigg\}.
\end{IEEEeqnarray*}
If $u_1 = u_2$ and there are corresponding
outputs $y_1 \neq y_2$, then
$z_R(u_1, u_2)$ is defined to be $\{\infty\}$.  If $R$ is single valued at $u_1$,
$z_R(u_1, u_1)$ is the empty set.

The \emph{Scaled Relative Graph} (SRG) of $R$ over $\mathcal{U} \subseteq \mathcal{H}$ is then given by
\begin{IEEEeqnarray*}{rCl}
        \srg[\mathcal{U}]{R} \coloneqq \bigcup_{u_1, u_2 \in\, \mathcal{U}}  z_R(u_1, u_2).
\end{IEEEeqnarray*}
If $\mathcal{U} = \mathcal{H}$, we write $\srg{R} \coloneqq \srg[\mathcal{H}]{R}$.
The SRG of a class of operators is defined to be the union of their individual SRGs.
Some examples of SRGs are shown in Figure~\ref{fig:composition}, (a), (b) and (c).

\subsection{Interconnections}

The power of SRGs lies in the elegant interconnection theory of 
\cite{Ryu2021}.  Given the SRGs of two systems, the SRG of their interconnection can
be bounded using simple graphical rules.  Given two systems $R$ and $S$, and subject
to mild conditions, we have:
\begin{IEEEeqnarray*}{rCl}
        \srg{\alpha R} &=& \srg{R \alpha} = \alpha \srg{R}\\
        \srg{R + S} &\subseteq& \srg{R} + \srg{S}\\
        \srg{R S} &\subseteq& \srg{R}\srg{S}\\
        \srg{R^{-1}} &=& \srg{R}^{-1}.
\end{IEEEeqnarray*}
For the precise meanings of these operations, and the requisite conditions on the
systems $R$ and $S$, we refer the reader to \citep{Ryu2021}.

\subsection{SRGs of systems}

The SRGs of LTI transfer functions are closely related to the Nyquist diagram, and
the SRGs of static nonlinearities are closely related to the incremental circle.
These connections are explored in detail in \citep{Chaffey2021c,
Pates2021}; below we
recall the two main results.  The h-convex hull is the regular convex hull with
straight lines replaced by arcs with centre on the real axis -- for a precise
treatment, we refer the reader to \citep{Huang2020a}.

\begin{thm}\label{thm:nyq}
        Let $g: L_2(\C) \to L_2(\C)$ be linear and time invariant, with transfer
        function $G(s)$.  Then
        $\srg{g} \cap \C_{\Im \geq 0}$ is the h-convex hull of $\nyq{G} \cap \C_{\Im \geq 0}$.
\end{thm}

\begin{thm}\label{prop:static_bound}
        Suppose $S: L_2 \to L_2$ is the operator given by a SISO static nonlinearity
        $s: \R \to \R$, such that for all $u_1, u_2 \in \R$, $y_i \in s(u_i)$, 
        \begin{IEEEeqnarray}{rCl}
        \mu (u_1 - u_2)^2 \leq (y_1 - y_2)(u_1 - u_2) \leq \lambda 
        (u_1 - u_2)^2.\label{eq:incremental_sector} 
\end{IEEEeqnarray}
        Then the SRG of $S$ is contained within the disc centred at 
        $(\mu + \lambda)/2$ with radius $(\mu - \lambda)/2$.        
\end{thm}

\begin{figure}[h!]
        \centering
        \includegraphics{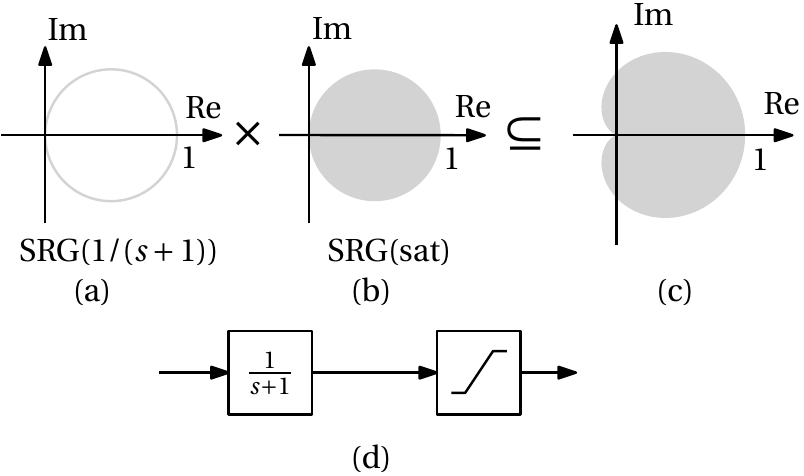}
        \caption{Bounding SRG for the composition of a first order lag and
        saturation.}%
                \label{fig:composition}
\end{figure}

Theorems~\ref{thm:nyq} and~\ref{prop:static_bound}, and the SRG interconnection
rules, allow us to construct bounding SRGs for arbitrary interconnections of LTI and
static nonlinear components.  A simple example is illustrated in
Figure~\ref{fig:composition}.

\section{Incremental robustness and sensitivity}

\subsection{Stability and incremental gain}
\begin{figure}[ht]
        \centering
        \includegraphics{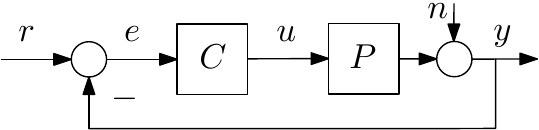}
        \caption{Negative feedback control structure.}%
        \label{fig:sym_fb}
\end{figure}

Given the negative feedback interconnection of Figure~\ref{fig:sym_fb}, incremental stability is guaranteed by the separation of 
the SRGs of $P^{-1}$ and $-C$, and the distance between them is an incremental
robustness margin, the reciprocal of which bounds the incremental gain of the
feedback system.  This is formalized in \citep[Thm. 2]{Chaffey2021c}; we recall the result
here.

Let $\mathcal{H}$ be a class of operators. By $\bar{\mathcal{H}}$, we will denote a
class of operators such that $\mathcal{H} \subseteq \bar{\mathcal{H}}$ and
$\srg{\bar{\mathcal{H}}}$ satisfies the chord property: 
if $z_1, z_2 \in \srg{\bar{\mathcal{H}}}$, then $\theta z_1 + (1-\theta)z_2 \in \srg{\bar{\mathcal{H}}}$ 
for all $\theta \in [0, 1]$.

\begin{thm}\label{thm:stability}
        Consider the feedback interconnection shown in Figure~\ref{fig:sym_fb}
        between any pair of operators $C \in \mathcal{C}$ and $P \in
        \mathcal{P}$, where $\mathcal{C}$ and $\mathcal{P}$ are classes of operators on $L_2$ with
        finite incremental gain.  If, for all $\tau \in (0, 1]$,
        \begin{IEEEeqnarray*}{rCl}
                \srg{\mathcal{C}}^{-1}\cap -\tau \srg{\bar{\mathcal{P}}} =
                \varnothing,
        \end{IEEEeqnarray*}
        then the incremental $L_2$ gain from $r$ to $u$ is bounded by $1/r_m$, where 
        $r_m$ is the shortest distance between
$\srg{\mathcal{C}^{-1}}$ and $-\srg{\bar{\mathcal{P}}}$.
\end{thm}

\begin{figure}[ht]
        \centering
        \includegraphics{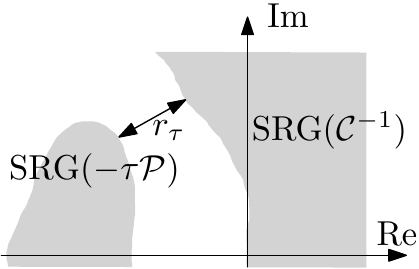}
        \caption{Illustration of Theorem~\ref{thm:stability}.}%
        \label{fig:margin}
\end{figure}

\subsection{The sensitivity SRG}

The operator $(I + PC)^{-1}$ maps $r$ to $e$ in the feedback system of
Figure~\ref{fig:sym_fb}, and the operator $(I - PC(-I))^{-1}$
maps $n$ to $y$.  These two operators have the same SRG, which we denote by
$\mathcal{S}$ -- the \emph{sensitivity SRG}.  

\begin{defn}
        The \emph{peak incremental sensitivity} is the maximum incremental gain of
        the operator $(I + PC)^{-1}$.
\end{defn}

The peak incremental sensitivity is equal to the maximum modulus of $\mathcal{S}$.
The following theorem gives the peak incremental sensitivity an interpretation as a
robustness margin.

\begin{thm}\label{thm:sensitivity}
        Let $s_m$ be the shortest distance between\\ $\srg{PC}$ and the point $-1$.  Then
        the peak incremental sensitivity is equal to $1/s_m$.
\end{thm}

\section{Loop shaping}

We demonstrate SRG loop shaping with two simple design examples for the control
structure shown in Figure~\ref{fig:sym_fb}. 
\subsection{Shaping for stability and robustness}

As first design example, we show how to use SRGs to ensure incremental stability of 
a closed loop system. Unlike traditional loop shaping, where the return ratio $L =
PC$ is modified, we graphically shape the inverse of the feedback system, $(P +
C^{-1})$, to 
improve the robustness of the closed loop. The use of SRGs makes the design close 
to classical Nyquist analysis, despite the nonlinearity of $P$.

\begin{figure}[ht]
        \centering
        \includegraphics{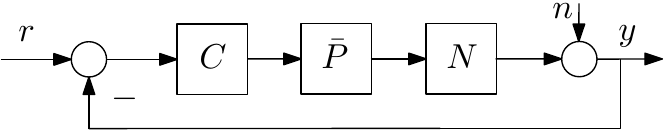}
        \caption{Example control system.  $\bar{P} = 1/(s(s+1))$, $N$ is a nonlinear
        operator and $C$ is the controller, to be designed. $r$ is the reference input, $n$
represents sensor noise.}%
\label{fig:fb2}
\end{figure}

Consider the system in Figure~\ref{fig:fb2}.  $C$ represents the controller, to be
        designed.  Suppose that the process consists of $\bar{P}$
        with LTI dynamics $1/(s(s+1))$, and a nonlinear operator 
        $N$, whose SRG is known to be bounded in the region illustrated in
        Figure~\ref{fig:composition} (c). We denote $C\bar{P}$ by $L$. The controller $C$ is to be designed to stabilize the system
and decrease the incremental gain.

\begin{figure}[ht]
        \centering
        \includegraphics[width=0.45\textwidth]{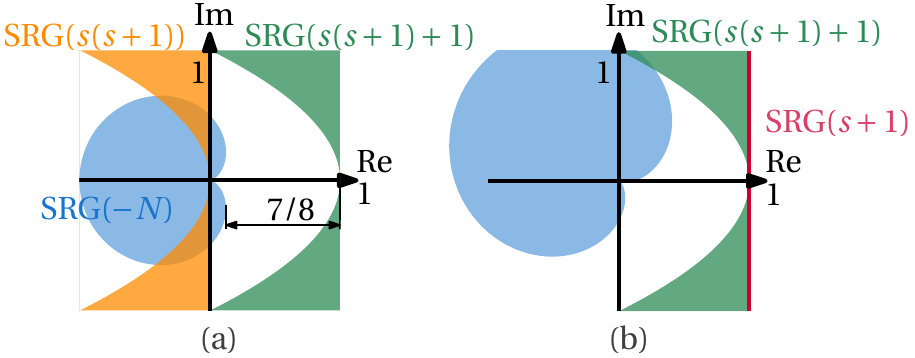}
        \caption{(a) SRGs of $-N$, $\bar{P}^{-1}$ and $(\bar{P} C_1)^{-1}$. (b) SRGs
        of $(\bar{P} C_1)^{-1}$, $(\bar{P} C_2)^{-1}$ and a scaled and rotated
nonlinearity, showing the improved robustness with $C_2$.}%
        \label{fig:example1}
\end{figure}

To ensure stability, we require the SRGs of $L^{-1} = (C\bar{P})^{-1}$ and $-N$ to be
separated, for all scalings of $N$ between $0$ and $1$ (following
Theorem~\ref{thm:stability}).  With $C_0 = 1$, the closed loop is unstable, as shown in
Figure~\ref{fig:example1} (a).
Shifting $L^{-1}$ to the left, by designing $C$ to give $L^{-1} = s(s+1) + 1$, gives a
stabilizing control. The controller reads $C_1 = s(s+1)/(1 + s(s+1))$.

To improve robustness and reduce the incremental gain of the system, the separation
of $\srg{L^{-1}}$ and $\srg{-N}$ must be increased (again, following
Theorem~\ref{thm:stability}).  For example, setting $L^{-1} = s + 1$ ($C_2 = s$) gives good
separation, and an incremental gain bound from $r$ to $u$ of $8/7 \approx 1.14$.  As
the incremental gain of $N$ is bounded by $1$ (the maximum modulus of its SRG), this
value also bounds the incremental gain from $r$ to $y$.  The increased separation of
the SRGs makes the system robust to uncertainties in the nonlinearity $N$, as
illustrated in Figure~\ref{fig:example1} (b).

\subsection{Shaping for performance}

We now focus on graphical methods for improving performance, and explore how the
sensitivity SRG can be shaped over particular sets of signals.
We consider a new system, again of the form of Figure~\ref{fig:sym_fb}, with $C = 1/(ks + 1)$, where $k$ is 
a scalar to be designed, and $P$ is a unit saturation.  The SRGs of $C$ and $P$ are
shown in Figure~\ref{fig:composition} (a) and (b).
        
Tracking performance and noise rejection are both characterized by the sensitivity
SRG.
Suppose that
we would like this SRG to have a
low modulus (corresponding to incremental gain) for signals with a bandwidth of $\omega_0 = 10$ rad$/$s and a maximum
amplitude of $2$.  The aim is to limit the maximum amplification of $(I +
PC)^{-1}$ over this range of signals.

A heuristic method is to maximize the distance between $\srg{PC}$ and $-1$
over the frequency range $[-\omega_0, \omega_0]$ and amplitude range $[-2, 2]$,
following Theorem~\ref{thm:sensitivity}.  This corresponds to maximizing the
minimum incremental gain of the inverse of the sensitivity operator over this range
of signals.

\begin{figure}[ht]
        \centering
        \includegraphics{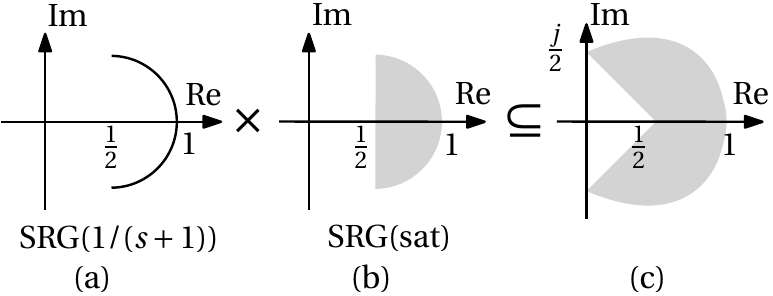}
        \caption{Left: SRG of $1/(ks + 1)$ over signals with bandwidth $[-1/k,
        1/k]$.  Right: SRG of saturation over signals with maximum amplitude $2$.}%
        \label{fig:example2}
\end{figure}

$\srg{PC}$ is bounded by the Minkowski product of $\srg{P}$ and $\srg{C}$.  
 Plotting the
 SRG of the saturation $P$ over the amplitude range $[-2, 2]$ gives the half-disc shown in
 Figure~\ref{fig:example2} (b).  The SRG of $C$ over $[-\omega_0, \omega_0]$ is described by 
\begin{IEEEeqnarray*}{rCl}
        \left( \frac{1}{1 + k^2 \omega^2}, j \frac{-k\omega}{1 + k^2 \omega^2}\right)
\end{IEEEeqnarray*}
for $\omega \in [-\omega_0, \omega_0]$.  As a first design, we can set $k$ so that
the bandlimited SRG of $C$ is half the circle (Figure~\ref{fig:example2} (a)) -- this is achieved by setting $k = 0.01$.  This gives the bound on $\srg{PC}$ shown in Figure~\ref{fig:example2} (c).
The minimum distance to the point $-1$ is $s_m = \sqrt{3}$. 

This method is, however, only a heuristic.  The saturation introduces higher
harmonics, so the assumption that signals have a bounded spectrum is invalidated when
the loop is closed.  However, given the lowpass properties of the system, the
approximation is reasonable.  The higher order harmonics of the output of the
saturation have low magnitude, and the unit lag has a lowpass behavior.  The
stability of the closed loop guarantees that these high frequencies are indeed
attenuated by the feedback system.  This assumption is similar to the lowpass
assumption of describing function analysis \cite{Slotine1991}.  The method here
differs from describing function analysis, however, in that arbitrary differences of bandlimited
inputs are considered, not just pure sinusoids.

\section{Other types of systems}

A significant advantage of the SRG is being able to place disparate system types on
an equal footing.  Like continuous time LTI systems, finite dimensional linear
operators described by matrices lend themselves well to shaping.  
\cite{Pates2021} has shown that the SRG of a matrix is equal to the numerical range
of a closely related, transformed matrix.  In the case of normal matrices, the SRG is
the h-convex hull of the spectrum \citep{Huang2020a}.  These results pave the way for
shaping a matrix's SRG by matrix multiplication and addition.

\begin{figure}[hb]
        \centering
        \includegraphics{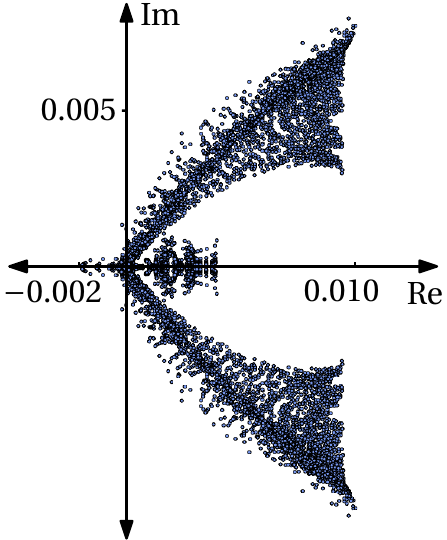}
        \caption{Sampling of the SRG of a potassium conductance.}
        \label{fig:K_srg}
\end{figure}

In cases where the analytic SRG is not available, the SRG can be sampled over the
signals of interest, and loop shaping methods can then be applied using the sampled
SRG.  For example, Figure~\ref{fig:K_srg} shows a sampled SRG of the potassium
conductance of the Hodgkin-Huxley model of a neuron \citep{Hodgkin1952}. 

\bibliography{srg}

\begin{thebibliography}{12}
\providecommand{\natexlab}[1]{#1}
\providecommand{\url}[1]{\texttt{#1}}
\providecommand{\urlprefix}{URL }
\expandafter\ifx\csname urlstyle\endcsname\relax
  \providecommand{\doi}[1]{doi:\discretionary{}{}{}#1}\else
  \providecommand{\doi}{doi:\discretionary{}{}{}\begingroup
  \urlstyle{rm}\Url}\fi

\bibitem[{Bode(1960)}]{Bode1960}
Bode, H.W. (1960).
\newblock Feedback -- the history of an idea.
\newblock In \emph{Proceedings of the Symposium on Active Networks and Feedback
  Systems}, Microwave {{Research Institute}} Symposia Series. {Polytechnic
  Press}, {Brooklyn}.

\bibitem[{Chaffey(2022)}]{Chaffey2021d}
Chaffey, T. (2022).
\newblock A rolled-off passivity theorem.
\newblock \emph{Systems \& Control Letters}, 162.

\bibitem[{Chaffey et~al.(2021)Chaffey, Forni, and Sepulchre}]{Chaffey2021c}
Chaffey, T., Forni, F., and Sepulchre, R. (2021).
\newblock Graphical nonlinear system analysis.
\newblock \emph{arXiv:2107.11272 [cs, eess, math]}.

\bibitem[{Clegg(1958)}]{Clegg1958}
Clegg, J.C. (1958).
\newblock A nonlinear integrator for servomechanisms.
\newblock \emph{Transactions of the American Institute of Electrical Engineers,
  Part II: Applications and Industry}, 77(1), 41--42.
\newblock \doi{10.1109/TAI.1958.6367399}.

\bibitem[{Hodgkin and Huxley(1952)}]{Hodgkin1952}
Hodgkin, A.L. and Huxley, A.F. (1952).
\newblock A quantitative description of membrane current and its application to
  conduction and excitation in nerve.
\newblock \emph{The Journal of Physiology}, 117(4), 500--544.
\newblock \doi{10.1113/jphysiol.1952.sp004764}.

\bibitem[{Huang et~al.(2020)Huang, Ryu, and Yin}]{Huang2020a}
Huang, X., Ryu, E.K., and Yin, W. (2020).
\newblock Scaled relative graph of normal matrices.
\newblock \emph{arXiv:2001.02061 [cs, math]}.

\bibitem[{McFarlane and Glover(1992)}]{McFarlane1992}
McFarlane, D. and Glover, K. (1992).
\newblock A loop-shaping design procedure using {{H}}/sub infinity / synthesis.
\newblock \emph{IEEE Transactions on Automatic Control}, 37(6), 759--769.
\newblock \doi{10.1109/9.256330}.

\bibitem[{Pates(2021)}]{Pates2021}
Pates, R. (2021).
\newblock The scaled relative graph of a linear operator.
\newblock \emph{arXiv:2106.05650 [math]}.

\bibitem[{Ryu et~al.(2021)Ryu, Hannah, and Yin}]{Ryu2021}
Ryu, E.K., Hannah, R., and Yin, W. (2021).
\newblock Scaled relative graphs: Nonexpansive operators via {{2D Euclidean}}
  geometry.
\newblock \emph{Mathematical Programming}.
\newblock \doi{10.1007/s10107-021-01639-w}.

\bibitem[{Slotine and Li(1991)}]{Slotine1991}
Slotine, J.J.E. and Li, W. (1991).
\newblock \emph{Applied Nonlinear Control}.
\newblock {Prentice Hall}, {Englewood Cliffs, N.J}.

\bibitem[{Vinnicombe(2000)}]{Vinnicombe2000}
Vinnicombe, G. (2000).
\newblock \emph{Uncertainty and Feedback: $H_\infty$ Loop-Shaping and the
  $\nu$-Gap Metric}.
\newblock Imperial College Press.

\bibitem[{Zames(1981)}]{Zames1981a}
Zames, G. (1981).
\newblock Feedback and optimal sensitivity: {{Model}} reference
  transformations, multiplicative seminorms, and approximate inverses.
\newblock \emph{IEEE Transactions on Automatic Control}, 26(2), 301--320.
\newblock \doi{10.1109/TAC.1981.1102603}.

\end{thebibliography}
\end{document}